\documentclass[useAMS,usenatbib]{mn2e}
\usepackage{times}
\usepackage{dcolumn}
\usepackage{graphicx}
\usepackage{aas_macros}
\usepackage{xspace}
\usepackage{amssymb}
\usepackage{ulem}

\newcommand{\msun}{\mbox{${\rm M}_{\odot}$}\xspace}
\newcommand{\myr}{\mbox {~${\rm M_{\odot}~\rm yr^{-1}}$}}
\newcommand{\y}{$\surd$}
\newcommand{\n}{$\times$}
\newcolumntype{d}{D{.}{.}{-1}}

\def\apgt{{\raise-.5ex\hbox{$\buildrel>\over\sim$}}\ }
\def\aplt{{\raise-.5ex\hbox{$\buildrel<\over\sim$}}\ }

\def\rev#1{{#1}}

\title[Chemical composition of ultra-compact binaries]{The chemical
  composition of donors in AM CVn stars and ultra-compact X-ray
  binaries: observational tests of their formation}

\author[Nelemans, Yungelson, van der Sluys \& Tout]{G.
  Nelemans$^{1}$\thanks{E-mail: nelemans@astro.ru.nl},
  L. R. Yungelson$^{2,1}$, M. V. van der Sluys$^{3}$ and Christopher A. Tout$^{4}$
  \\
  $^{1}$Department of Astrophysics, IMAPP, Radboud University
  Nijmegen, P.O. Box 9010, NL-6500 GL Nijmegen, The Netherlands\\
  $^{2}$Institute of Astronomy of the Russian Academy of
  Sciences, 48 Pyatnitskaya Str., 119017 Moscow, Russia \\
  $^{3}$Department of Astrophysics, Northwestern University, USA, now at
  Department of Physics, University of Alberta, Edmonton, Canada\\
  $^{4}$Institute of Astronomy, University of Cambridge, Madingley Road, Cambridge CB3 0HA, UK\\
}

\begin{document}

\date{Accepted . Received \today}

\pagerange{\pageref{firstpage}--\pageref{lastpage}} \pubyear{2007}

\maketitle

\label{firstpage}

\begin{abstract}
We study the formation of ultra-compact binaries (AM CVn stars and
ultra-compact X-ray binaries) with emphasis on the surface chemical
abundances of the donors in these systems.  Hydrogen is not
convincingly detected in the spectra of any these systems.  Three
different proposed formation scenarios involve different donor stars,
white dwarfs, helium stars or evolved main-sequence stars.  Using
detailed evolutionary calculations we show that the abundances of
helium white dwarf donors and evolved main-sequence stars are close to
equilibrium CNO-processed material, and the detailed abundances
correlate with the core temperature and thus mass of the
\textit{main-sequence} progenitors.  Evolved main-sequence donors
typically have traces of H left.  For hybrid or carbon/oxygen white
dwarf donors, the carbon and oxygen abundances depend on the
temperature of the helium burning and thus on the helium core mass of
the progenitors. For helium star donors in addition to their mass, the
abundances depend strongly on the amount of helium burnt before mass
transfer starts and can range from unprocessed and thus almost equal
to CNO-processed matter, to strongly processed and thus C/O rich and
N-deficient.  We briefly discuss the relative frequency of these cases
for helium star donors, based on population synthesis
results. \rev{Finally we give diagnostics for applying our results to
observed systems and find that the most important test is the N/C
ratio, which can indicate the formation scenario as well as, in some
cases, the mass of the progenitor of the donor. In addition, if
observed, the N/O, O/He and O/C ratios can distinguish between helium
star and white dwarf donors.} Applied to the known systems we find
evidence for white dwarf donors in the AM CVn systems GP Com, CE~315
and SDSS~J0804+16 and evidence for hybrid white dwarf or very evolved
helium star donors in the ultra-compact X-ray binaries 4U~1626-67 and
4U~0614+09.
\end{abstract}


\section{Introduction}
\label{sec:intro}

AM CVn stars and ultra-compact X-ray binaries (UCXBs) are interacting
double stars with orbital periods less than about one hour, with white
dwarf or neutron star accretors \citep[e.g.][]{vh95,nel06}. This
distinguishing property implies that the orbits are so tight, that
only compact, evolved donors, such as helium stars, white dwarfs or
low-mass stars with hydrogen-deficient envelopes, fit in. Indeed the
optical spectra of these systems lack any convincing sign of hydrogen
but instead show helium lines in absorption or emission in the AM CVn
systems \citep[e.g.][]{1995Ap&SS.225..249W}, or weak C/O or He/N lines
in emission in the UCXBs
\citep{njm+04,wnr+06,njs06,2008A&A...485..183I}. For one of the UCXBs,
4U~1626-67 \citet{sch+01} have found double peaked O and Ne lines in
the X-ray spectrum. The short orbital periods and close proximity of
AM CVn stars and UCXBs make them the brightest Galactic gravitational
wave sources \citep[e.g.][]{nyp03,rgn+06}.

In recent years many new ultra-compact binaries have been discovered
\citep[e.g.][]{rgm+05,ahh+05,2008AJ....135.2108A,2009MNRAS.394..367R,2005ApJ...634L.105D,bji+06,zjm+07},
bringing the total number of known AM CVn stars to 22 and the known
number of (candidate) UCXBs to 27.

\begin{figure}
  \includegraphics[angle=0,width=\columnwidth,clip]{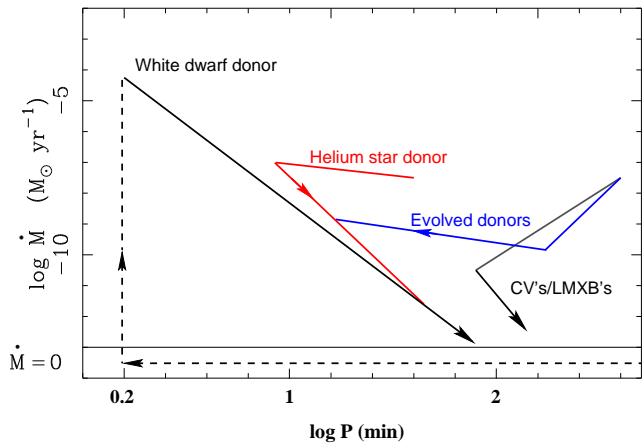}
  \caption{Sketch of the period -- mass transfer rate evolution of the
    binaries in the three proposed formation channels (the dashed line shows
    the detached phase of the white dwarf channel). For comparison, the
    evolutionary path of an ordinary hydrogen-rich CV or low-mas X-ray binary
    is shown.}
\label{fig:P_Mdot}
\end{figure}

Three formation channels have been proposed for the formation of
interacting ultra-compact binaries, schematically depicted in
Fig.~\ref{fig:P_Mdot} \citep[for more details see,
e.g. ][]{2005ASPC..330...27N,2006LRR.....9....6P}.

$\bullet~$ { \textit{\textbf{The white dwarf channel}}. The first
  channel is a compact binary of a low-mass white dwarf with a compact
  companion (higher-mass white dwarf or neutron star) with an orbital
  period short enough that angular momentum loss by gravitational wave
  radiation drives the stars into contact within the age of the
  Universe \citep[e.g.][\rev{see
  Fig.~\ref{fig:P_Mdot}}]{pw75,ty79a,nrs81}. The low-mass white dwarf
  donor fills its Roche lobe and starts mass transfer to the
  companion. For a sufficiently low mass ratio of the components, the
  system may enter stable mass transfer and evolve to longer periods
  \citep[see][ and references therein]{ynh02,mns02}.

 $\bullet~$ \textit{ \textbf{The helium star channel}}. Alternatively, the
  direct progenitor of the donor star may be a helium core burning star that
  overfills its Roche lobe, transferring mass to a white dwarf or neutron star
  \citep{skh86,1987ApJ...313..727I,tf89,ef90,ty96}.  The ensuing mass transfer
  is stable if the helium star is not much more massive than the accreting
  star. The binary evolves to shorter periods owing to angular momentum loss
  by gravitational wave radiation \rev{at typical mass-transfer rates of a few
    $10^{-8} \myr$ (see Fig.~\ref{fig:heabund} below).} This stage of
  evolution lasts for $10^6 - 10^7\,$yr.  At a some point the shortening of
  the period \rev{is turned around.  This is due to a change in the
    mass--radius relation of the donor as it becomes degenerate and its
    chemical composition changes.  The star begins to grow as it loses mass
    and the orbit must expand to accommodate it.  Note that mass transfer from
    a less to a more massive star generally expands the orbit and now the
    mass-transfer rate adjusts so that this expansion sufficiently dominates
    the shrinkage driven by angular momentum loss} \citep[for more details
    see][]{yun08}}.  The orbital period $P_{\rm orb}$ attains a minimum, which
is typically close to 10\,min and the system then evolves to longer $P_{\rm
  orb}$.  After several hundred million years the donors become homogeneous
degenerate objects and the helium star and white dwarf families of
ultra-compact binaries become indistinguishable \citep{dtw+08,yun08}.  Because
of the lifetimes of the various phases of evolution the overwhelming majority
of ultra-compact binaries formed via the helium-star channel should be seen in
the post-$P_{\rm orb, min}$.

$\bullet~$ \textit{ \textbf{The evolved main-sequence star channel}}.
  The third channel starts with a main-sequence star transferring mass
  to a white dwarf or neutron star.  This is a cataclysmic variable
  (CV) or low-mass X-ray binary (LMXB). If the main-sequence star
  begins mass transfer sufficiently late on the main sequence and if
  sufficient angular momentum is lost, by for instance magnetic
  braking, the hydrogen deficient core of the donor can be exposed
  during mass transfer so that the star evolves to far below the usual
  period minimum for population-I CVs \citep[of about
  $80$\,min;][]{ps81,2002PASP..114.1108T,2005ApJ...635.1263W,2009MNRAS.397.2170G}.  In the
  most extreme cases they can fall to a period minimum of around
  $10\,$min
  \citep[e.g.][]{1985SvAL...11...52T,1986ApJ...304..231N,tfe+87,2003MNRAS.340.1214P,2003ApJ...598..431N}.

One of the problems to distinguish these different formation
channels is that they more or less lead to the same donors, very
low-mass degenerate dwarfs.  Studies of the population of objects in
the phase \textit{before} they become ultra-compact binaries could
give insight into the origin of ultra-compact binaries.  However,
currently the putative progenitors are not observed in large enough
numbers for such a study.

An alternative is to study the details of the chemical composition of
the donor-stars \citep{nt02}, which show up in the optical or X-ray
spectra of these systems as discussed above, or alternatively
for UCXBs in the properties of the thermonuclear explosions (type I
X-ray bursts) that occur on the surface of the accreting neutron stars
\citep[e.g.][]{bil95,btd+03,icv05,2003ApJ...595.1077C}.


In this paper we model the chemical composition of the donor stars in
ultra-compact binaries in detail to find out whether this can distinguish
between the different proposed formation channels. In Section~\ref{results} we
present the detailed calculations of the donor abundances and in
Section~\ref{population} we investigate the distribution of abundances that is
expected for the different formation channels. In Section~\ref{diagnostics} we
develop a set of diagnostics that can be used to distinguish the formation
channels given observed abundances or abundance ratios or limits thereon and
apply these results to observed systems in Section~\ref{application}. We
summarise and discuss our results in Section~\ref{sec:discon}.

\section{Abundance patterns in the donors of ultra-compact binaries}\label{results}

For all calculations we used the Eggleton stellar evolution code TWIN
\citep[][ Eggleton 2006, private
communication]{egg71,egg72,pte+95,2002ApJ...575..461E}, with opacity
tables taken from OPAL \citep{1992ApJ...397..717I} and
\citet{1994ApJ...437..879A}.  Convection is modelled by mixing length theory
\citep{bohmvitense1958} with the ratio of mixing length to pressure scale height
$l/H_P=2.0$.  Mixing is modelled by a diffusion
equation for the abundances.
Zero-age main-sequence stars have solar metallicity
($X=0.70$, $Z=0.02$).  The code explicitly follows the abundances of H, He,
C, N, O, $^{22}$Ne and Mg and uses the the pp-chain and CNO bi-cycle for
hydrogen burning and $3\alpha$ and ${\rm
^{12}C(\alpha,\gamma)^{16}O}$ reactions for helium burning.
 
We model angular-momentum loss by magnetic braking according to
equation~(34) of \citet{1983ApJ...275..713R} with $\gamma=4$.  Following
\citet{prp02}, we reduce the strength of the magnetic braking by an
ad-hoc factor $\exp\left(1-0.02/q_\mathrm{conv}\right)$, where
$q_\mathrm{conv}$ is the mass fraction of the convective envelope of
the donor star and assume that magnetic braking abruptly shuts off
when $q_\mathrm{conv} = 1$.  Angular-momentum loss owing to
gravitational-wave emission is taken into account with a standard
prescription \citep[e.g.][]{1975ctf..book.....L}.  

\subsection{The white dwarf channel}\label{res:wd}

\begin{figure}
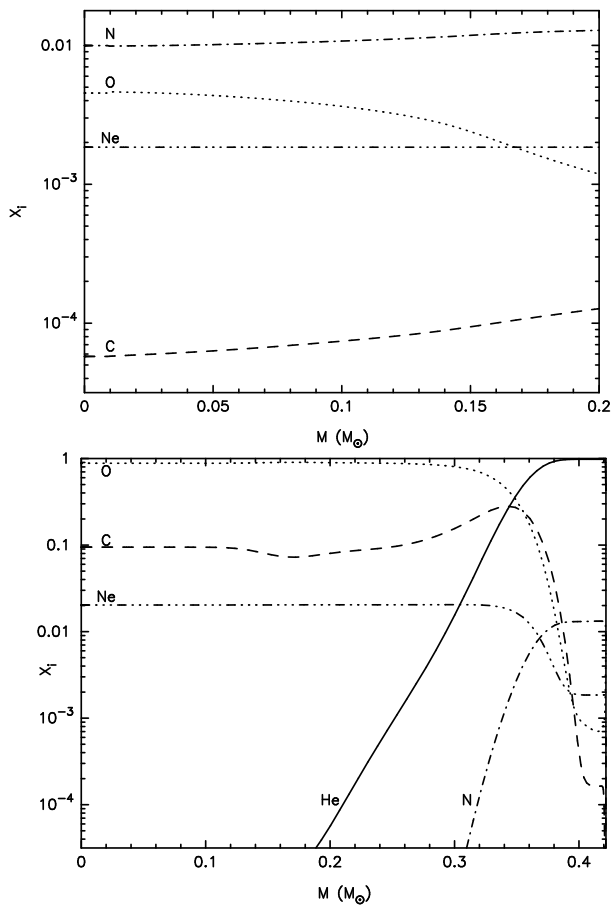

\includegraphics[angle=-90,width=0.95\columnwidth,clip]{inter_02hewd.eps}
\hspace*{0.3cm}
\includegraphics[angle=-90,width=0.9\columnwidth,clip]{inter_0421hybwd.eps}
  \caption{Abundances in a \rev{0.2 \msun} helium white dwarf donor
    (top) and a \rev{0.421 \msun} hybrid white dwarf donor (bottom) as function of
    internal mass coordinate. The abundance of He in the helium white
    dwarf is 0.98.  For the hybrid dwarf we do not plot hydrogen-rich
    outermost $0.006\,\msun$.  The helium white dwarf is the
    descendant of a $1\,\msun$ star, the hybrid white dwarf descends
    from a $3\,\msun$ star. }
\label{fig:wdabund}
\end{figure}

For the AM CVn stars the donors are expected to be helium white dwarfs, as
heavier white dwarf donors would either lead to unstable mass transfer and a complete
merger of the system or to mass-transfer rates far in excess of the Eddington
limit for the accreting white dwarf \citep{npv+00,mns02}. The most
massive donors are expected to be no more massive than about $0.3\,\msun$.
They are born with hydrogen-helium envelopes of about $0.01\,\msun$.  For
UCXBs similar mass-transfer stability arguments have led us to conclude that
the donors must have masses less than about $0.45\,\msun$ \citep{ynh02} allowing
helium white dwarfs and hybrid white dwarfs as donors.  Hybrid WDs
have C/O cores and thick (about $0.1\,\msun$) He-C-O mantles and some H in the
outermost layers (see Fig.~\ref{fig:wdabund}). Hybrid white dwarfs are the
remnants of low- or intermediate-mass stars that overflow their Roche lobes prior
to non-degenerate helium ignition in the core
\citep{it85,2000MNRAS.319..215H,2002MNRAS.335..948C,2003MNRAS.341..662C,2007MNRAS.382..779P}.

In Fig.~\ref{fig:wdabund} we show the abundances of He, C, N and O for a
helium and a hybrid white dwarf as function of internal mass coordinate.  The
composition of the helium white dwarf is obviously dominated by helium with
CNO abundances according to the equilibrium of the CNO cycle while that of the
hybrid white dwarf donors is dominated by O and C because the helium mantle is
very quickly lost, in the first few Myr after Roche-lobe overflow (RLOF), at
very short orbital periods (less than about $15\,$min). The interior parts of the
stars which are of interest ($M_r \lesssim 0.1\,\msun$) are virtually
homogeneous so we do not have to do the actual binary evolution calculations
of the white dwarf donors but can simply map the internal structure of
representative donor stars to the expected abundances as function of period,
just as we did in \citet{nt02}.  In doing so, we used the mass--radius relation for
zero-temperature white dwarfs \citep{vr88} and neglected the typically small
effects of finite entropy of the donors
\citep[e.g.][]{bil02,db03,dbn05,dtw+08}. The donor stars have an outer
convective zone that gradually penetrates inward \citep{dtw+08}.  This
changes the composition of the transferred material but again
only at very short periods, at which hardly any systems are observed.


\begin{figure}
\includegraphics[angle=-90,width=0.95\columnwidth]{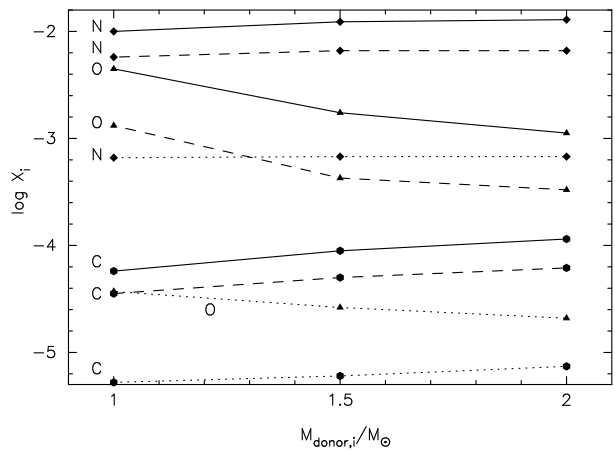}
  \caption{Abundances in the centres of helium white dwarf donors as
function of their progenitor mass and metallicity. Solid lines connect
markers for $Z=0.02$, dashed lines -- for $Z=0.01$, dotted lines -- for
$Z=0.001$.}
\label{fig:abund_HeWD1}
\end{figure}

The equilibrium CNO abundances depend on the temperature of the stellar core
\citep[e.g.][]{1999A&A...347..572A} and thus differ for different mass
main-sequence progenitors of the helium white dwarfs. We therefore calculated
the inner structure of helium white dwarfs descending from progenitors with
masses of 1, 1.5 and $2\,\msun$.  In Fig.~\ref{fig:abund_HeWD1} we show the
resulting central abundances as function of the progenitor mass and
metallicity.  As Fig.~\ref{fig:wdabund} shows, these abundances are well
representative for the entire interior of white dwarf \rev{and because the
  central abundances do not change during the growth of the helium core, they
  are the same for all helium white dwarfs descending from the same
  progenitor}. The plot shows that the abundance ratios, the distances between
the lines, which are easier to determine observationally, differ for different
progenitor masses and hence can in principle be used to constrain these.
Metallicity also plays a role but, for the remainder of this article, we focus
on Solar metallicity because we are mainly concerned with relatively nearby
objects \rev{that predominantly come from the thin disk}.

For the hybrid white dwarfs the C and O abundances are expected to be mainly a
function of the mass of the helium core and not the details of the preceding
evolution. We computed a set of evolutionary sequences for binaries with
primary masses of 2.5 and $3\,\msun$ and compared the resulting hybrid white
dwarfs with hybrid white dwarfs formed as a result of evolution of helium
stars \citep[][Section~\ref{res:he}]{yun08}.  Indeed, the abundances for similar
helium core/star masses are quite similar.  We use a grid of hybrid white
dwarfs in the range $0.35 < M/\msun < 0.65$.

\subsection{The helium star channel}\label{res:he}

Detailed evolutionary calculations of binaries consisting of a low-mass helium
star and a white dwarf are presented in \citet{yun08} and we refer the reader
to that paper.  Initial models of helium stars were constructed
from a $0.46\,\msun$ core of a star with ZAMS mass of $3.16\,\msun$. Initial
abundances of species were $Y=0.98$, ${\rm X_C=0.00017}$, ${\rm X_N=0.013}$,
${\rm X_O=0.00073}$, ${\rm X_{Ne}=0.00185}$, ${\rm X_{Mg}=0.00068}$. 
\citet{yun08} assumed \rev{no mass loss from the binary} for systems with
initial helium star masses of 0.35, 0.4 and $0.65\,\msun$. However, thermonuclear
explosions at the surface of the accreting white dwarfs may cause the mass
transfer to be non-conservative
\citep[e.g. ][]{1987ApJ...313..727I,it91,2004A&A...419..645Y,2007ApJ...662L..95B}.
A repeat of the calculations for completely non-conservative evolution
(the lost matter takes away specific angular momentum of the
accretor, i.e. we mimick in this way helium nova explosions) yielded
virtually identical abundances as function of orbital period. We also ran two
sets of calculations, completely conservative and completely
non-conservative, for the case of a neutron star accretor with initial
helium donor masses of 0.35, 0.65 and $1\,\msun$ and again found no
significant difference in the abundances as function of orbital period.

Because the onset of mass transfer from the helium star \rev{strongly
suppresses} the helium burning \citep[e.g.][\rev{figures 3 and
4}]{skh86}, the chemical composition of the core of the donor depends
sensitively on the \textit{moment} at which the helium star fills its
Roche lobe.  We use evolutionary calculations for binaries that start
mass transfer almost immediately after the common-envelope phase in
which the helium star is formed, as well as systems that fill their
Roche lobe just before core helium exhaustion. In this way the
complete range of expected abundances can be probed, although we would
like to note that the extremes of this range will likely be rare in
practice, because the periods need to be fine tuned. We discuss this
in more detail in Section~\ref{population}.

\begin{figure*}
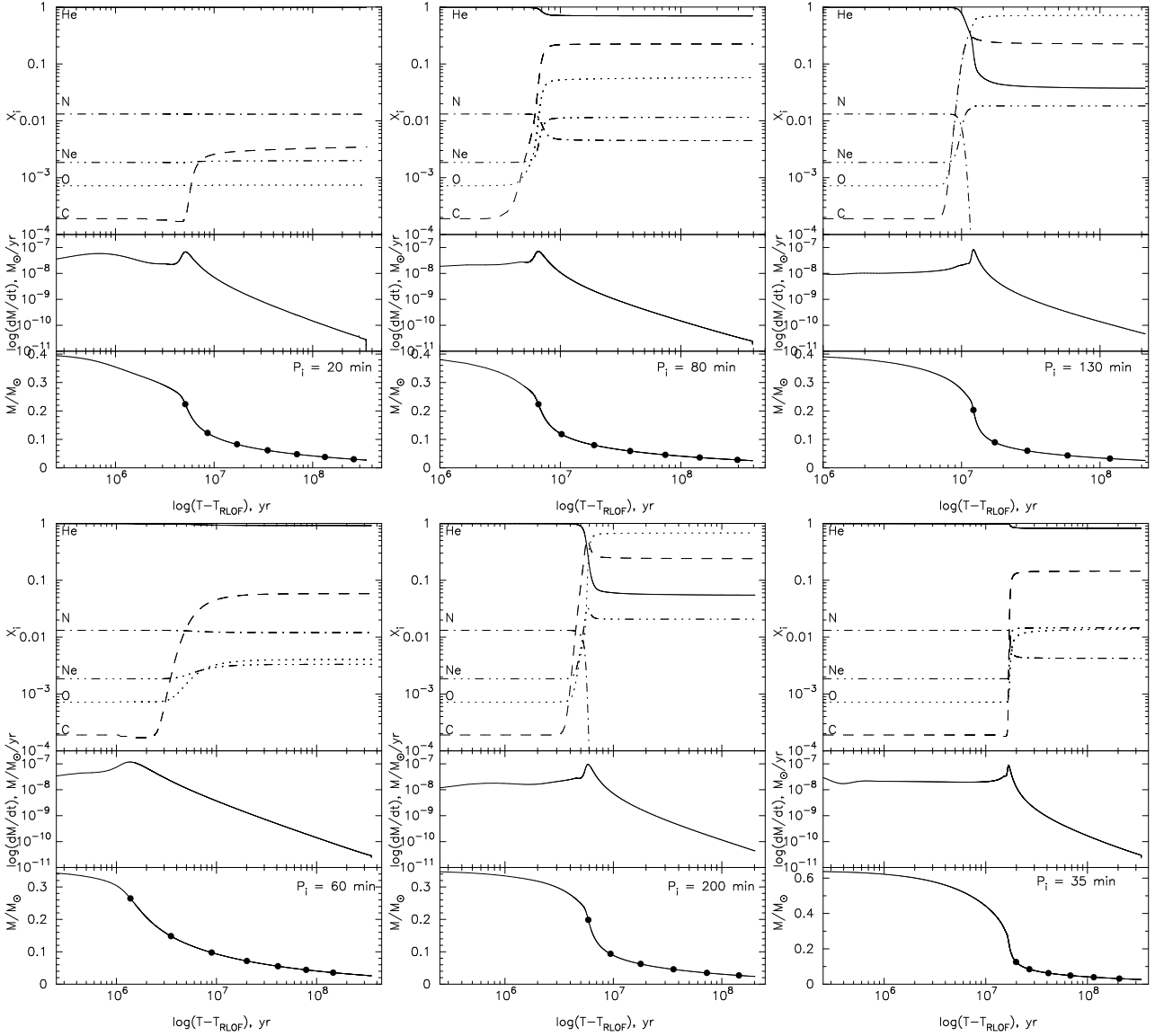

\includegraphics[width=5.6cm,clip]{tot0040620.eps}
\includegraphics[width=5.6cm,clip]{tot0040680.eps}
\includegraphics[width=5.6cm,clip]{tot00406130.eps}

\includegraphics[width=5.6cm,clip]{tot0351460.eps}
\includegraphics[width=5.6cm,clip]{tot03514200.eps}
\includegraphics[width=5.6cm,clip]{tot0651435.eps}
 \caption{Overview of the evolution and abundances for helium star
   donors in ultra-compact binaries, showing \rev{surface} abundances
   (top), mass-transfer rate (middle) and donor mass (bottom) as a
   function of time since the start of Roche-lobe overflow. The binary
   period is indicated by the solid circles in the bottom panels for
   periods of $ P_{\rm orb} = P_{\rm orb, min} $, 15, 20, 25, 30, 35
   and 40 min (left to right; not all tracks reach beyond 30 min). The
   initial periods are indicated in the bottom panels. Top row are
   sequences with white dwarf accretors, in the order of increasing
   initial orbital period and thus amount of He-processing before
   RLOF. The bottom row is for neutron star accretors, with two
   different initial helium star masses. }
\label{fig:heabund}
\end{figure*}

In Fig.~\ref{fig:heabund} we show a representative set of evolutionary
sequences for binaries with helium donors and white dwarf/neutron star
accretors.  The full set of sequences is published on-line.  For each sequence
we plot the \rev{surface} abundances of the transferred material, the mass
transfer rate, donor mass and orbital period as function of the time since the
onset of RLOF. This allows full assessment of the evolution, both in the
initial phase when the binary evolves to shorter periods at almost constant
mass transfer rate of a few times $10^{-8}\,\myr$, as well as later,
when the system has passed its period minimum.  Before the period minimum
stars lose matter that was outside the convective core in the helium-burning
stage so the material is helium rich, with CNO abundances corresponding to the
CNO cycle equilibrium for relatively massive stars that are typical
progenitors of helium stars\footnote{It has been proposed that many of the
  subdwarf B stars in the Galaxy are produced by low-mass stars with
  degenerate cores that experience a helium flash just after losing their
  hydrogen envelope in an enhanced stellar wind or binary interaction
  \citep[e.g.][]{ddr+96,hpm+02}. These helium stars would have different
  equilibrium CNO abundances, more like the helium white dwarfs. However, in
  our population estimates (Section~\ref{population}) we find that these
  systems are typically too wide for RLOF during core helium burning.}.
Shortly after the period minimum (or for the most evolved donors already
before period minimum) helium burning products, most notably C and later O,
come to the surface.  Depending on the initial period of the binary the
enrichment by C and O can be very mild, or C and O can dominate even He.

In the top row of Fig.~\ref{fig:heabund} we plot sequences for an
initially $0.4\,\msun$ helium star donor transferring material to an
initially $0.6\,\msun$ white dwarf accretor. The leftmost plot is for
the case of a post-common-envelope period of 20 min, in which RLOF
starts almost immediately after formation of the helium star and very
little helium burning occurs, so He, N, O and $^{22}$Ne abundances are
virtually unchanged.  However, after the period minimum the carbon
abundance increases substantially. For an intermediate
post-common-envelope period of 80\,min, there is a dramatic change of
abundances around period minimum because the layers which were in the
core and experienced some He-burning are exposed and for most of the
AM CVn evolution C and O (and even $^{22}$Ne) dominate over N. The He
abundance is noticeably reduced. Finally for the most evolved donor,
with initial period of 130 min (helium abundance in the core at RLOF
$Y_{\rm c} \approx 0.07$), the changes are even more dramatic and O
and C dominate even over He.  Nitrogen becomes extinguished even
before $ P_{\rm orb, min} $.  Note that in the most extreme cases
He-burning continues for some time after RLOF but He is still not
completely burnt so significant amounts of He should still be
detectable, contrary to hybrid white dwarf donors.

In the bottom row of Fig.~\ref{fig:heabund} a number of evolutionary sequences
for systems with neutron star accretors are shown. The leftmost panel is again
for fairly short initial period of 60\,min and shows quite a change in C but
not much in the other elements. The middle plot is for the sequence with the
longest initial period for which RLOF starts when the star almost totally
burnt helium in its core ($ Y_{\rm c} \approx 0.06 $) and looks very similar
to the most evolved donor shown for the sequence with a white dwarf accretor.
The bottom right plot is for a He star plus neutron star system with initial
donor mass of $0.65\,\msun$ and initial orbital period close to the minimum
possible for such a system, 35\,min It shows that at a given orbital period
there may be a scatter of a factor of several in abundance ratios, depending
on the initial mass of the donor.

\begin{figure}
\includegraphics[width=0.9\columnwidth,clip]{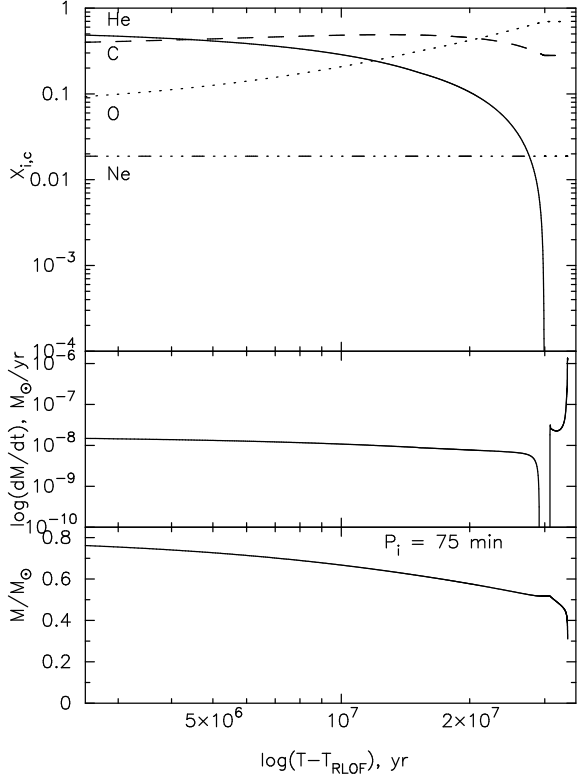}
\caption{Evolution of a helium star companion with $M_0=0.8\,\msun$ to a
neutron star in a system with initial period of 75\,min as a function of
time since the start of Roche-lobe overflow.  Top is the abundances in the
stellar core, middle is the mass-transfer rate and bottom is the donor mass.  }
\label{fig:evolHe}
\end{figure}

 A peculiar evolutionary path is followed by initially relatively massive
 helium stars (more than about $0.65\,\msun$) with neutron star
companions.  These
 overfill their Roche lobes after burning of a substantial fraction of helium
 in the core and continue He-burning during the semidetached stage of
 evolution.  For instance, the system with $M_{\rm He}=0.80\,\msun$ and $
 P_0=70$\,min starts mass loss when $Y_c \approx 0.643$ and proceeds along a
 conventional evolutionary track. In a slightly wider initially system
 with $ P_0=75$\,min, RLOF occurs when $Y_c \approx 0.56$ (see
 Fig.~\ref{fig:evolHe}).  In this system the donor detaches from its Roche
 lobe when its mass has decreased to $0.52\,\msun$ and $Y_c \approx 0.006$. The
 orbit continues to shrink and mass exchange resumes in the helium-shell
 burning stage. However, when the donor mass is more than about $0.45\,\msun$, mass loss
 cannot be stabilised by mass and angular-momentum loss from the system
 \citep[by isotropic re-emission, see][]{ynh02} and the ensuing
 mass loss proceeds on a dynamical time scale (see
 Fig.~\ref{fig:evolHe}). Thus, such a system does not contribute to the helium
 star channel for UCXBs.  Evolutionary sequences for $1\,\msun$ donor stars
 follow a similar path irrespective of the amount of He burnt prior to RLOF.
 The details of this type of evolution will be discussed in a forthcoming
 paper (Yungelson et al. in preparation).\footnote{For systems with white
   dwarf accretors such an evolutionary path was encountered for a
$(0.65+0.8)\,\msun$ system in which the He-star had $Y_c \approx 0.19$ at RLOF
   \citep{yun08}.}  Thus, there are two factors limiting the helium star
   channel for the formation of ultra-compact binaries, the maximum post-common-envelope
   period for which mass transfer still starts during core helium burning and
   a limiting mass above which the system detaches as described above (see
   also Fig.~\ref{fig:hechannel_MP}).

\subsection{The evolved main-sequence star channel}\label{res:evolved}

\begin{figure}
\includegraphics[angle=-90,width=\columnwidth,clip]{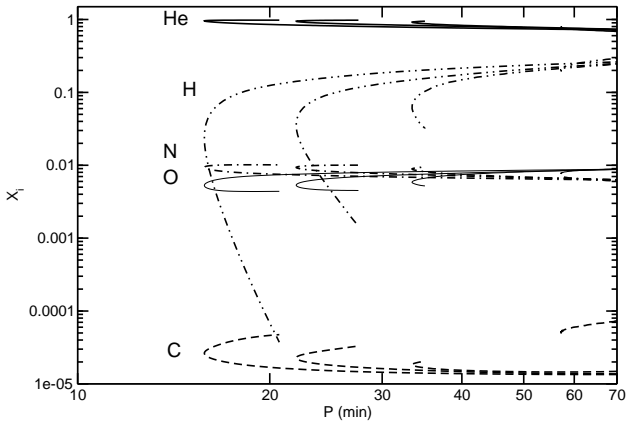}
 \caption{Surface abundances of
evolved main-sequence donors as function of orbital period for the
phase of evolution close to the period minimum.  The track with the
shortest period minimum and the one that has a period minimum just
below 70 minutes as well as two intermediate cases are shown.}
\label{fig:pre_min}
\end{figure}

\begin{figure}
\includegraphics[angle=-90,width=\columnwidth,clip]{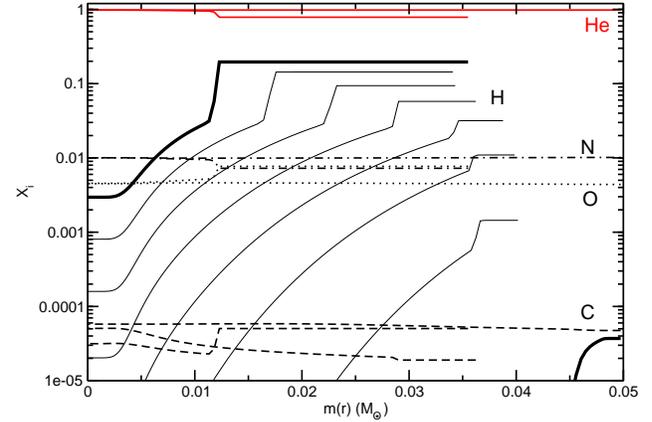}
 \caption{Internal chemical structure of the last models of the tracks
with evolved main-sequence donors. For the He, C, N and O abundances
only the two (three) extreme tracks are shown, all other tracks fall
within that region.  The H abundance is shown for all tracks, ranging
from the least evolved, that has a period minimum just below 70 min
(top heavy solid line), to the most evolved system (heavy solid line
just visible in the right bottom corner). }
\label{fig:structure}
\end{figure}

The formation of ultra-compact binaries from main-sequence donors
  requires two conditions.  First, the initial periods are such that
  the progenitors fill their Roche lobe close to the end of the main
  sequence.  Second, angular momentum loss drives the components
  together at a sufficient rate that the ultra-short periods can be
  reached within the Hubble time
  \citep{1985SvAL...11...52T,tfe+87,2003MNRAS.340.1214P,2005A&A...431..647V}.

\rev{We focus here on ultra-compact binaries that have lost (almost) all of
  their hydrogen\footnote{We will briefly discuss the few cataclysmic
    variables with periods below the period minimum in Sect.~\ref{sec:evCV}},
  using as an example, results of a detailed calculation for a system with a
  $1.0\,\msun$ white-dwarf accretor and a $1.0\,\msun$ donor, taken from a
  grid computations (Van der Sluys et al. in preparation).  We follow
  \citet{2005A&A...431..647V} and assume all the mass accreted by a white
  dwarf is expelled from the system by subsequent nova explosions, taking with
  it specific angular momentum of the accretor. This is justified by the fact
  that mass loss from the donor is always below $10^{-8}\,\myr$,
  the approximate upper limit of accretion rates at which strong nova
  explosions occur \citep[e.g.][]{yaron05}. Systems with initial abundances of
  $X=0.7$, $Z=0.02$ and initial periods between 2.43 and~$2.89\,$d reach a
  period minimum below 70 min. within $13\,$Gyr\footnote{This period limit is
    appropriate because the longest currently known period of an AM~CVn star
    (CE~315) is $65.1\,$min.}}.

\rev{In most cases the period minimum is only mildly shorter than
  70\,min. This is because unless the donor is very evolved, it
  becomes completely mixed and for $X \apgt 0.4$ the minimum period is $\apgt
  70$\,min. \citep[][consistent with this work]{1986ApJ...304..231N}.
  In Fig.~\ref{fig:pre_min} we show the resulting surface abundances
  of four tracks that avoid mixing and thus evolve to shorter
  period. The H abundance decreases because the inward penetration of
  the outer convection zone, but before period minimum never drops
  below 0.01. He is enhanced and C, N and O evolve towards equilibrium,
  but typically have $\mathrm{N/O} \approx 2$, while in helium white
  dwarf descendants of more massive stars N/O is close to 10.}

\begin{figure*}
\includegraphics[angle=-90,width=0.9\columnwidth,clip]{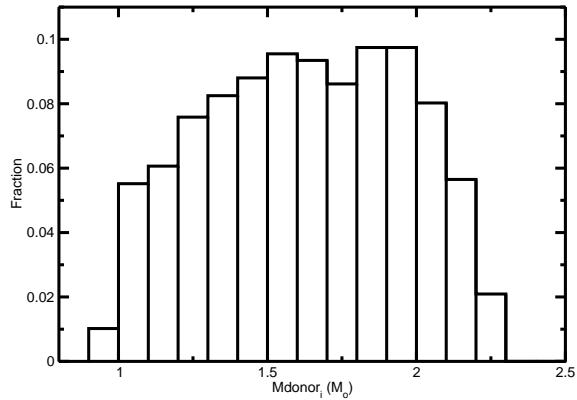}
\hspace*{0.5cm}
\includegraphics[angle=-90,width=0.9\columnwidth,clip]{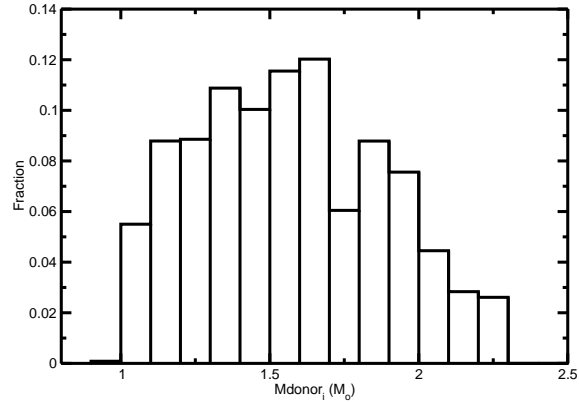}
 \caption{Progenitor masses of the helium white dwarf donors in AM CVn stars
   (left) and UCXBs (right). }
\label{fig:hewd_progenitors}
\end{figure*}

\rev{We were not able to evolve the models far past period minimum, but the
  further abundance evolution can be illustrated by looking at the internal
  structure of the last models (Fig.~\ref{fig:structure}). The surface
  convective region penetrates deeper and deeper until the donors become fully
  mixed, in agreement with other calculations
  \citep{1985SvAL...11...52T,tfe+87,1989Ap&SS.151..125F,2003ApJ...598..431N}.
  We cannot follow the balance of losing less processed material on the
  outside and progressive mixing, but the limiting abundances are set by the
  current outer and inner abundances. As can be seen in
  Fig.~\ref{fig:structure} the amount of H that is still present in the donors
  after the period minimum depends strongly on the initial period and may
  change significantly owing to mixing. We cannot at the moment say more than
  that the H abundances can be anywhere from 0.2 down to below $10^{-5}$, with
  only the upper limit possibly affected by mixing. In more extreme cases than
  presented above, when the donor already has a tiny helium core at RLOF, the
  H abundance may be zero \citep{1989Ap&SS.151..125F,2003ApJ...598..431N} but
  for a $1\,\msun$ star this may require more than a Hubble time of evolution.
  In any case, such systems should be exceptionally rare. The abundances of
  elements other than hydrogen do not vary much and are similar to those of
  normal helium white dwarfs descended from $1\,\msun$ stars but may have
  slightly less He, N and~C and slightly more O.}
 
Thus the donors in this channel still have hydrogen, although it can be at the
fraction of per cent level. However, H lines in the optical spectra are so
easily formed that even for number ratios $\mathrm{H/He} \approx 1/100$
hydrogen should be still detectable
\citep[e.g.][]{1982ApJ...257..672W,mhr91,2009A&A...499..773N}.

\section{Expected progenitor population: what abundances do we expect?}
\label{population}

In order to get an idea of the expected abundance patterns typical for
the different formation channels we have a look at the progenitors of the
donors indicated by population synthesis calculations.  We caution here
that the parameters entering calculations are still not well restricted and
these results should be interpreted with care.  Most of the results presented
here for AM CVn stars are from the calculations of \citet{nyp03}.  They used an
angular momentum balance formalism for the description of unstable mass
transfer between components of comparable masses \citep{nvy+00} and a time
and position dependent star-formation history based on the Galaxy model of
\citet{bp99}.

For UCXBs we consider only systems in which neutron stars formed
through core-collapse following the formation of an iron core. That means we
do not consider the possibility of formation of neutron stars via
accretion-induced collapses of white dwarfs
\citep[e.g. ][]{nomkondo91}. In our assumptions we followed the
modelling of the population of Galactic binary neutron stars
\citep{py99} with updates presented for example by \citet{lyhnp05}. We note
that \citet{bt04} find that accretion induced collapse of a massive
ONe white dwarf in an ultra-compact binary is the dominant formation
channel for UCXBs.  They find similar numbers of helium (60\%)
and hybrid (40\%) white dwarf donors and very few helium star
donors. However, the relative fractions of systems formed via
different channels are very sensitive to the population
synthesis parameters. For UCXBs, the crucial parameters are the
common-envelope efficiency and the efficiency of accretion on to white
dwarfs. We will address these issues in the forthcoming study.

\subsection{White dwarf channel}

\begin{figure}
\includegraphics[angle=-90,width=0.9\columnwidth,clip]{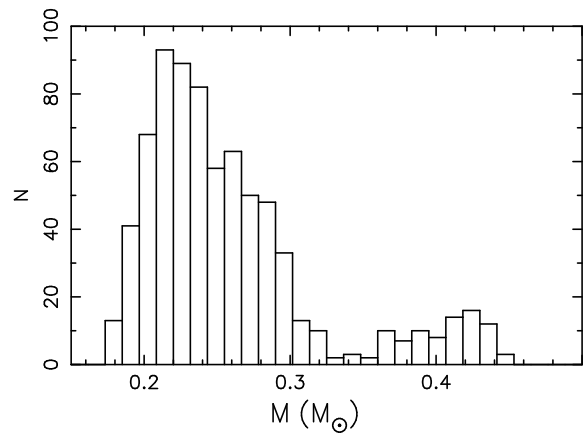}
\caption{Histogram of the initial mass distribution of white dwarf
  donors in UCXBs. The majority are helium white dwarfs, with masses
  below  about $0.35\,\msun$, with a smaller contribution of hybrid white
  dwarfs with slightly higher masses.}
\label{fig:UCXB_donors}
\end{figure}

As discussed in Section~\ref{res:wd}, for the AM CVn systems we expect
only helium white dwarf donors in this channel, so that any differences in the abundance
patterns should be due to the main-sequence mass of the progenitor. In
Fig.~\ref{fig:hewd_progenitors} (left panel) we show a histogram of
the expected progenitor main-sequence masses. The distribution is
fairly flat
between 1 and $2\,\msun$, so that no strong bias for the abundance patterns is
expected. A similar plot but for the helium white dwarf donors in
UCXBs is shown in the right panel of
Fig.~\ref{fig:hewd_progenitors}. Again, no strong bias towards either
low or high masses is expected. For UCXBs, in addition to helium white
dwarfs, hybrid white dwarf
donors also are allowed from mass transfer stability arguments
\citep{ynh02}. In Fig.~\ref{fig:UCXB_donors} we show the distribution
of initial white dwarf masses of the donors in UCXBs. The low-mass
helium white dwarfs dominate the expected population. A
detailed comparison of the population of UCXBs with the observed
systems is deferred to a
forthcoming study of that issue.

\subsection{Helium star channel}


\begin{figure*}
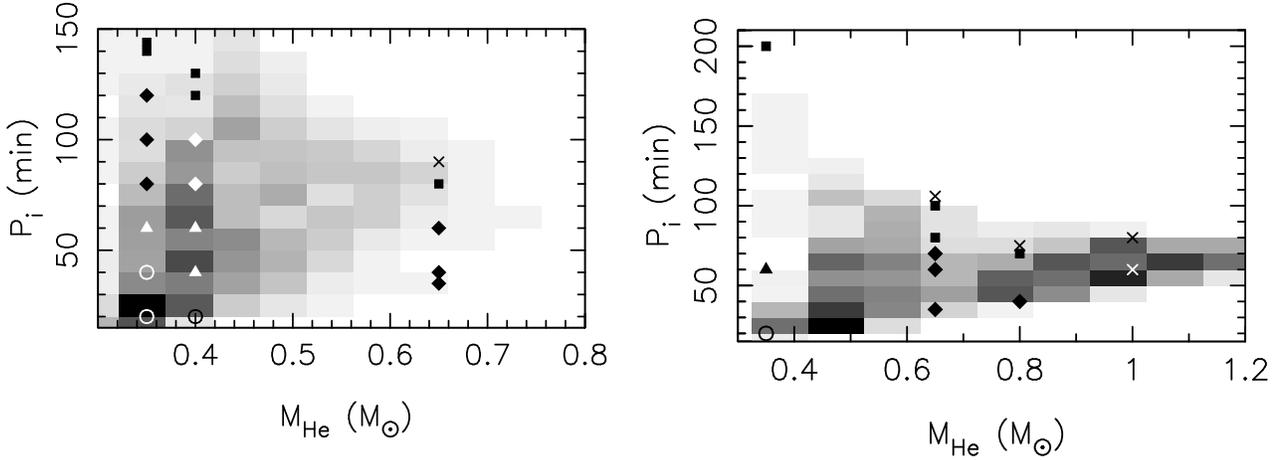

\includegraphics[angle=-90,width=0.95\columnwidth,clip]{hechannel_MP_new.ps}
\hspace*{0.3cm}
\includegraphics[angle=-90,width=0.99\columnwidth,clip]{hechannel_UCXB_new.ps}
\caption[]{Distribution of helium star masses and orbital periods for
  binaries forming AM CVn stars (left) and UCXBs (right) via the
  helium star channel. Over-plotted are the different classes of
  abundance patterns ranging from almost completely unprocessed
  (circles, $X_{\rm C}, X_{\rm O} < X_{\rm N} < X_{\rm He}$), via two intermediate steps
  (triangles $X_{\rm O} < X_{\rm N} < X_{\rm C} < X_{\rm He}$,
diamonds $X_{\rm N} < X_{\rm C}, X_{\rm O} <
  X_{\rm He}$) to strongly processed material (squares $X_{\rm N} <
X_{\rm He} < X_{\rm C}$ or $X_{\rm O}$).  Crosses mark initial parameters of the systems
  in which the donors complete He-burning during mass transfer, detach
  from their Roche lobes and resume mass transfer after evolving into
  hybrid white dwarfs. 
}.
\label{fig:hechannel_MP}
\end{figure*}

 For the helium star channel\footnote{We identify here ultra-compact
binaries with systems that passed through the period minimum and are
already homogeneous ($ P_{\rm orb} \gtrsim 20 $\,min, see
Fig. ~\ref{fig:heabund}).} the abundance patterns depend completely on
the initial post-common envelope period of the binary system, as longer initial periods
lead to later RLOF, when the helium star has burnt more of its helium
in the core. We classified the abundance patterns according to a
simple scheme based on the fact that  during helium burning,
first the C abundance starts to rise, until it becomes larger than the
N abundance, then the N abundance starts to drop and the O abundance
becomes larger, until N drops below both C and O. Finally the He
abundance starts to drop and first the C and then the O abundance
starts to dominate (see Fig.~\ref{fig:heabund}).  The categories we
have are

$\bullet$ unevolved ($X_{\rm C}, X_{\rm O} < X_{\rm N} < X_{\rm He}$) for systems with
initial periods very close to the periods of RLOF,

$\bullet$ two intermediate steps ($X_{\rm O} < X_{\rm N} < X_{\rm C} < X_{\rm He}$) and
($X_{\rm N} < X_{\rm C}, X_{\rm O} < X_{\rm He}$),

$\bullet$ and strongly processed ($X_{\rm N} < X_{\rm He} < X_{\rm C}$
or
$X_{\rm O}$) for systems that have almost exhausted helium before RLOF.

 In Fig.~\ref{fig:hechannel_MP} we show the expected distribution of
helium star masses and orbital periods of the progenitors of AM CVn
stars and UCXBs that form via the helium star channel.  The details of
this scenario for the formation of UCXBs will be discussed in our
forthcoming paper on the Galactic population of UCXBs.  The different
abundance patterns are over-plotted as the symbols.  Although there is
a concentration of systems towards lower masses and shorter periods, a
mixture of abundance patterns is expected from the helium star
channel.  We would like to stress here once again that while
abundances for given $ P_i$ and $M_{\rm He}$ are rather firmly set by
evolutionary computations, the distribution of underlying systems is a
sensitive function of input parameters of population synthesis.

\subsection{Evolved main-sequence star channel}

For the evolved main-sequence star channel, the main distinction in
abundances arise via the mass of the main-sequence star and the extent
of H-exhaustion at the moment of RLOF.
Extremely narrow ranges of initial masses and orbital periods which
can lead to periods shorter than about $70\,$min. in a Hubble time
\citep[see grids of models in][]{2003MNRAS.340.1214P,2005A&A...431..647V} lead to a
fairly small range of expected abundance ratios of CNO-group elements.
H/He in this channel may vary by two orders of magnitude (see
Figs.~\ref{fig:pre_min} and \ref{fig:structure}) and in extreme cases
may drop to zero, if H becomes undetectable.

\section{Diagnostics for determining formation channel and progenitor properties}\label{diagnostics}

As we have seen in the previous sections, different proposed
progenitors to ultra-compact binaries do have different chemical
compositions but in particular the helium star channel shows a lot of
diversity and overlap with the other channels. We here discuss a
possible diagnostic to use if abundances and in particular abundance
ratios or limits on them are available from the observations.
In general the abundances  and thus expected spectra are either
dominated by He and N, or by C and O. We therefore first select on
these two features.

\subsection{Sources showing N (and He)}

\begin{figure}
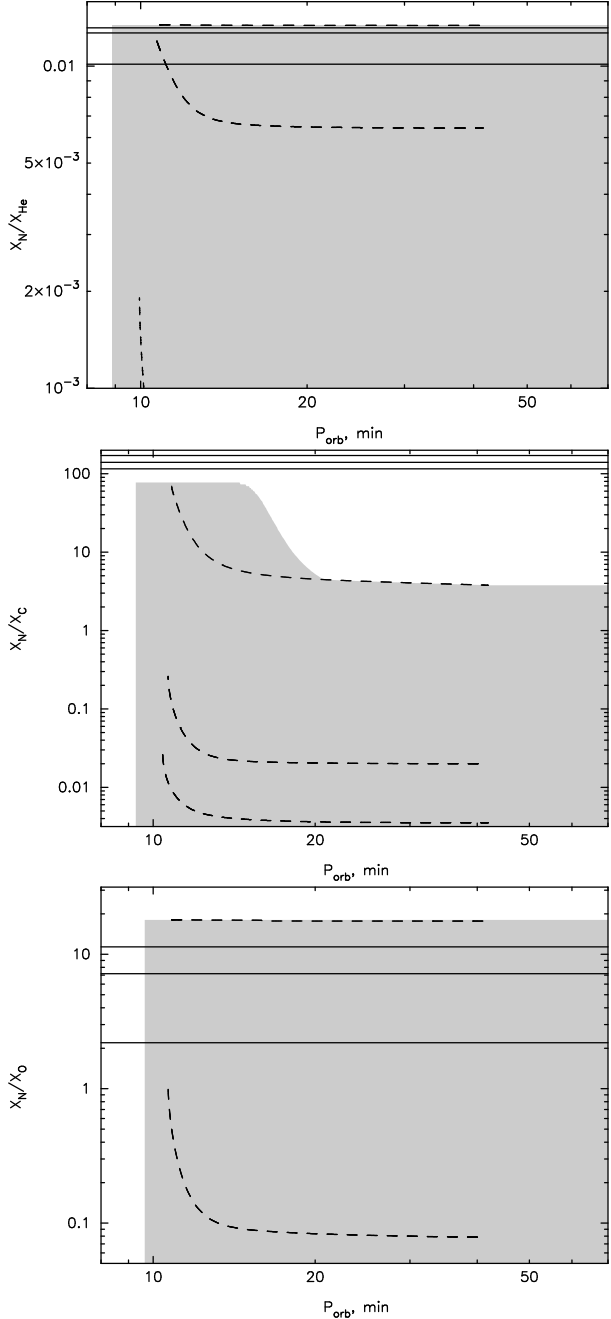

\includegraphics[angle=-90,width=0.95\columnwidth,clip]{nheratiowdhest1.eps}
\includegraphics[angle=-90,width=0.95\columnwidth,clip]{ncratiowdhest1.eps}
\includegraphics[angle=-90,width=0.95\columnwidth,clip]{noratiowdhest1.eps}
\caption{Abundance ratios (by mass) N/He, N/C and N/O for the helium white
  dwarf (solid line) and helium star (shaded regions with dashed lines) donors
  as a function of orbital period.  \rev{The shaded regions show the range
    covered by the many different helium star tracks, although the bottom of
    the range extends to 0.}  Dashed lines \rev{show the abundance ratios of
    the helium star donors shown in detail in the upper panel of
    Fig.~\ref{fig:heabund}}. The helium white dwarfs are descendants of 1, 1.5
  and $2\,\msun$ stars (top to bottom in two upper panels and bottom to top in
  the lower one). \rev{We plot the central abundance ratios, which should
    be a good representation for the whole period range and do not depend on
    the initial helium white dwarf donor mass}.  In particular the N/C ratio
  seems to be a good discriminator of the different formation channels.}
\label{fig:Nratios}
\end{figure}

Helium white dwarf, helium star and evolved main-sequence star donors
can be helium-rich with CNO abundances dominated by N. In
Fig.~\ref{fig:Nratios} we show the abundance ratios N/He, N/C and N/O
for helium white dwarfs and helium stars. The N/C ratios are well
separated, especially at periods above 20 min, with $\rm N/C < 10$ for all
helium star models.  This is true even for those that fill their Roche lobe virtually
immediately after exiting the common-envelope phase, because of
vigorous production of C by very moderate He-burning.  $\rm N/C > 100$ for all helium white dwarfs. Thus even non-detection of C may
constrain the formation channels, at least if the upper limits on C
are strong enough. If C is detected, the N/C ratio for helium
stars gives a good indication of the extent of helium exhaustion
before RLOF has started.

For the helium white dwarf donors, once the very high N/C ratio has
been determined and confusion with helium star donors is excluded, the
N/O ratio may actually put interesting constraints on the progenitor
mass of the helium white dwarf.  This ratio ranges from about~2 for
descendants of about $1\,\msun$ stars to more than about 10 for descendants of
about $2\,\msun$ stars (Fig.~\ref{fig:Nratios}, lower panel).

\begin{figure}
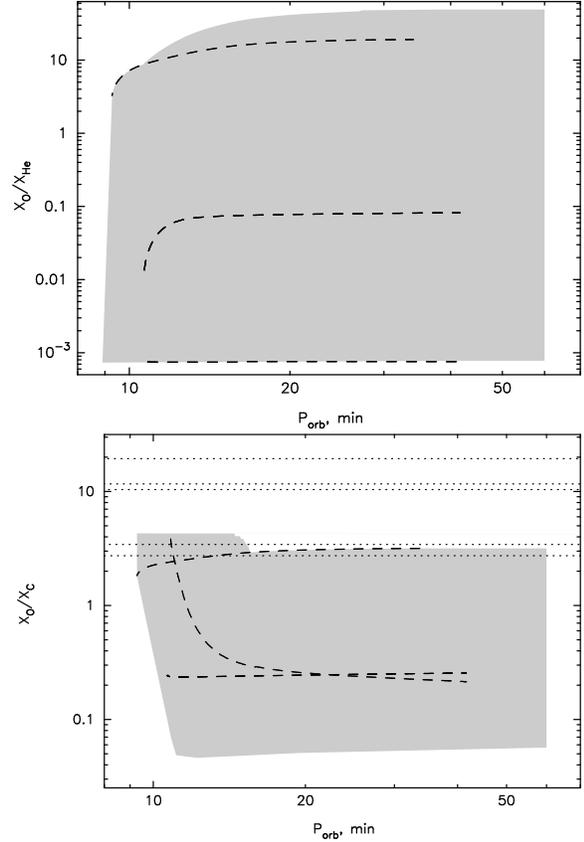

\includegraphics[angle=-90,width=0.9\columnwidth,clip]{oheratiowdhest.eps}
\includegraphics[angle=-90,width=0.9\columnwidth,clip]{ocratio2.eps}
\caption{Abundance ratios (by mass) O/He and O/C for the hybrid white dwarf
  (dotted lines) and helium star (shaded region and dashed lines) donors as
  function of orbital period. The dashed lines correspond to the tracks
  \rev{in the upper panel of} Fig.~\ref{fig:heabund}.  \rev{The hybrid white
    dwarfs have initial masses between 0.35 to 0.65 \msun and we show their
    central abundances. These should be a good representation for the whole
    period range.} The ranges of ratios for hybrids formed by case B mass
  exchange and hybrids that descend from helium stars overlap. Hybrid donors
  have lower O/C for higher mass. \rev{Note that He is absent in the interiors
    of hybrid white dwarfs.}}
\label{fig:Oratios}
\end{figure}

\subsection{Sources showing C or O}
When the donors have experienced significant helium burning, as for
some helium star donors and for hybrid white dwarf donors, the N may
 decrease to undetectable levels or be totally burnt but He, C
or O may show up. In Fig.~\ref{fig:Oratios} we show the expected O/He
and O/C ratios for helium star donors and hybrid white dwarf
donors. As expected there is no He left in the hybrid white dwarf
donors and the O/He ratio increases from less than $0.01$ to more than
$10$ with
increasing fraction of burnt helium in the helium star donors. The
most evolved helium star donors have similar O/C ratios as the most
massive hybrid white dwarf of about $0.65\,\msun$, with values around 3,
but the more relevant and lower mass hybrids have higher ratios,
around 10.

\section{Application to observed binaries}\label{application}

Optical and X-ray observations of ultra-compact binaries have provided
limits on the presence of certain elements, in particular, H, He, C,
N, O and $^{22}$Ne, Si, Ca and Fe. We separate the two classes in this
discussion on the application of our findings to the observed
binaries.  A caveat has to be made that it is not easy to
determine abundances from observed spectra, because of uncertainties
in the structure of accretion discs.

\subsection{AM CVn stars}

In the upper part of table~\ref{tab:abundances} we list the detected elements in AM CVn
stars. Observationally, their optical spectra fall in two categories,
emission line spectra for the longer period systems (plus ES Cet) and
absorption line spectra in the shorter period systems. The absorption
line systems (AM CVn and HP Lib and the outbursting systems in their
high states) typically hardly show any elements other than He, with
the exception of quite weak Mg and Si lines in AM CVn. More promising
are the emission line systems, in which N lines, in addition to the
very strong He lines, are always detected if the appropriate spectral
range is observed.  In particular GP~Com has been studied in detail, with
strong limits found on the presence of C and O from the absence of C
lines and weakness of the O lines in the optical spectra
\citep[][]{mhr91},
yielding estimates of $\rm N/C > 100$ and $\rm N/O \approx 50$.  
\citet{1995MNRAS.274..452M} report the discovery of C in the UV
spectrum of GP Com and, scaling the line flux ratios to those
observed in CVs, they estimate $\rm N/C \approx 10$. However, they comment on the
fact that this is incompatible with earlier optical results and
suggest the UV may underestimate the N abundance.
For CE~315 (V396
Hya), a similar conclusion can be drawn because the system is very similar to
GP Com \citep[see][]{2009A&A...499..773N}.  From UV spectra \citet{gsd+03} derived a flux ratio of
N\textsc{iv}/C\textsc{iv} $> 14.5$ in agreement with CNO
processing. For GP Com and CE~315 the puzzle is the absence of heavy
elements such as Mg, Si, Ca and Fe that would be expected.  This led
\citet{mhr91} to suggest a low metallicity for the progenitor and leaves
the question of where the abundant N has come from. In any case the high
N/C  (assuming the UV estimate in GP Com indeed is a lower limit)
points to a helium white dwarf donor (Fig.~\ref{fig:Nratios}), while
in that case the high N/O ratio points towards a progenitor of the
helium white dwarf on the high mass side.  A similar conclusion can be
drawn for the most recently discovered AM~CVn star (SDSS~J0804+16) for which N/C
$>$ 10 and N/O $\gtrsim$ 10 have been derived
\citep{2008arXiv0811.3974R}.

Some limits have been derived from the X-ray spectra of AM~CVn
stars. \citet{str04b} derives detailed abundances for GP Com of $X_{\rm He} = 0.99,
X_N = 1.7\, \times 10^{-2}, X_{\rm O} = 2.2\, \times 10^{-3}, X_{\rm
  Ne} = 3.7\, \times 10^{-3},X_{\rm S} = 2.3\,
\times 10^{-4}, X_{\rm Fe} = 8\, \times 10^{-5}$ and a limit on C $
X_{\rm C} < 2\, \times 10^{-3}$. This is
roughly consistent with the results from the optical spectra, except
for a lower
N/O ratio of about 8. \citet{rhm+05} find slightly higher values $
X_{\rm N} = 3\,
\times 10^{-2}, X_{\rm O} = 6\, \times 10^{-3}$, a ratio of $\rm N/O =
5$.  \citet{rhm+05,rgm+06} find, in
addition, evidence for enhanced N in AM CVn, HP Lib, CR Boo, SDSS~J1240-01 and
CE~315, as expected from the He-rich optical spectrum.

For the other AM CVn systems, the determination of the donor type is
not so clear.  Prominent N lines manifest themselves in the red
part of the spectrum and that is also where the strongest C
and O lines would show up, if they would be present. However, most
studies of AM CVn stars have been made in the blue part of the
spectrum. For SDSS~J1240-01 there is a red spectrum in which indeed
the N lines, as well as quite strong Si lines are detected but no
sign of C or O lines \citep{rgm+05}. Although no detailed calculations
have been made, a simple estimate, with the same LTE model that was
used by \citet{mhr91}, \citet{njm+04} and \citet{2008arXiv0811.3974R}
suggest that
$\rm N/C > 1$ without doubt.  This rules out hybrind white dwarf donors and helium star donors,
except the least evolved ones.

\subsection{Ultra-compact X-ray binaries}

In table~\ref{tab:abundances} we also list the detected elements in
UCXBs.  These come from X-ray, UV and optical spectroscopy, as well as
abundances inferred from the properties of type I X-ray
bursts. Unfortunately in many cases the detection of elements is
uncertain and, even in the cases where the detections are solid, it
is typically impossible to derive meaningful abundance ratios because
model spectra are not yet very realistic
\citep[e.g.][]{njm+04,wnr+06}.

 The extremely short orbital period of 4U~1820-30 led to the
suggestion that the transferred material in this system must be
hydrogen-deficient \citep[e.g. ][]{tfe+87,1988ApJ...324..851M}.  The
properties of type I X-ray bursts has led to the conclusion that the
transferred material in 4U~1820-30 is indeed helium, possibly with a
small amount of H \citep{bil95,2003ApJ...595.1077C}.  This is consistent
with an evolved main-sequence donor if H really is present, or with a
helium white dwarf or helium star donor if not. However,
\citet{2005A&A...431..647V} have shown that, even with full magnetic
braking,
this scenario can be discarded because it requires very finely tuned
initial parameters and predicts many systems with $10 <
P_\mathrm{orb}/{\rm min} < 60$ for each observed 10\,min binary.

For many other systems there are now optical spectra that show no
evidence for any H (or He in many cases), while all longer period
low-mass X-ray binaries always show strong H lines. However, the
accretion disc spectral models of \citet{wnr+06} suggest that amounts
up to 10\,per cent of H and He could remain undetected.

\begin{table*}
\caption[]{Detected elements in AM CVn stars and UCXBs.  \n ~marks the
  absence of certain element, while it should be detectable if
  present, \y ~detection of the element, ? ~possible detection, dash
  ~no information.}

\begin{tabular}{llllllllllll}
Object     & H  & He & C  & N  & O  & Ne & Mg & Si & Ca & Fe & Notes and references\\\hline
ES Cet     & \n & \y & ?  & ?  & -  & -  & -  & -  &  - & -  & Bowen blend, i.e. N and/or C (1)  \\
AM CVn     & \n & \y & -  & ?  & -  & -  & \y & \y & -  & -  & Estimate $X_N = 0.03$ from X-rays (2) \\
HP Lib     & \n & \y & -  & ?  & -  & -  & -  & -  & -  & -  & Estimate $X_N = 0.02$ from X-rays (2)  \\
CR Boo     & \n & \y & -  & ?  & -  & -  & -  & -  & -  & -  & Estimate $X_N = 0.015$ from X-rays (2) \\
V803 Cen   & \n & \y & -  & \y & -  & -  & -  & \y & -  & ?  & (3)                               \\
CP Eri     & \n & \y & -  & -  & -  & -  & -  & \y & -  & -  & (4)                               \\
2003aw     & \n & \y & -  & -  & -  & -  & -  & \y & \y & \y & (5)                               \\
SDSS~J1240-01   & \n & \y & \n & \y & \n & -  & -  & \y & - & \y & (6) Certainly N/C $>$ 1                \\
SDSS~J0804+16  & \n & \y & \n & \y & \n & -  & ?  & \y  & \y  & \y  & N/C $>$10, N/O $\gtrsim$10 (7)        \\
GP Com     & \n & \y & \n & \y & \n & \y & \n & \n & \n & \n & N/C $>$ 100, N/O $\gtrsim$5-10 (8-10, 2)\\
SDSS~J1552+32  & \n & \y & -  & -  & -  & -  & \y & -  & -  & -  & (11)                              \\
CE~315      & \n & \y & \n & \y & \n & -  & -\n& \n & \n & \n & (12, 25)                              \\ \hline
4U 1820-30 &  ? & ?  & -  &  - & -  & -  & -  & -  & -  & -  & He, possibly H [Type I X-ray burst (13,14)]\\
4U 1543-62 & \n & \n &\y? & \n &\y? & ?  & -  & -  & -  & -  & X-ray (15) and optical (16) spectra \\
M15 X-2    & -  & ?  & ?  &  - & -  & -  & -  & -  & -  & -  & UV spectrum (17)\\
4U 0614+09 & \n & \n & \y & \n & \y & ?  & -  & -  & -  & -  & X-ray (18) and optical (16,19) spectra\\
4U 1626-67 & \n & \n & \y & \n & \y &\y  & -  & \y & -  & -  & X-ray, optical and UV (20,21,19,22)\\
2S 0918-54 & \n & ?  &  ? &  - &  ? & ?  & -  & -  & -  & -  & X-ray bursts (23), optical spectra (16) \\
4U 1916-05 & \n & \y & \n & \y & \n & -  & -  & -  & -  & -  & Optical spectrum (21)    \\
XTE 0929-314& ?  & ?  & ?  & ?  & ?  & -  & -  & -  & -  & -  & Optical spectra (21)  \\
A 1246-58  & \n & ?  & ?  & -  & ?  &  ? & -  & -  & -  &  - & X-ray bursts and optical spectrum (24)  \\
\end{tabular}

\smallskip
References (1) Steeghs et al. in prep. (2) \citet{rhm+05} (3) \citet{rgn+07}, (4) \citet{gns+01}, (5) \citet{rgm+05c}, (6) \citet{rgm+05}, (7) \citet{2008arXiv0811.3974R}, (8) \citet{mhr91} (9) \citet{1995MNRAS.274..452M} (10) \citet{str04b}, (11) \citet{2007MNRAS.382.1643R}, (12) \citet{rrg+01}, (13) \citet{bil95} (14) \citet{2003ApJ...595.1077C} (15) \citet{jc03} (16) \citet{njm+04} (17) \citet{2005ApJ...634L.105D} (18) \citet{jpc00} (19) \citet{wnr+06} (20) see \citet{sch+01} (21) \citet{njs06} (22) \citet{haw+02} (23) \citet{icv05} (24) \citet{2008A&A...485..183I} (25) \citet{2009A&A...499..773N}
\label{tab:abundances}
\end{table*}

The best constraints are found for 4U~1626-67 and 4U~0614+09, for which
optical spectra exclude large amounts of H or He and show C and O
lines \citep{njm+04,wnr+06,njs06}. The X-ray spectrum of 4U~1626-67
shows double peaked O and Ne emission lines \citep{sch+01}, while its
UV spectrum shows strong C and O lines but not the usual He and N
lines. This all suggests evolved helium stars or hybrid white dwarfs
as donors for these two systems. \citet{njm+04} show that the optical
spectrum of 4U~1543-62 is similar to that of 4U~0614+09.  This suggests a
C/O rich donor too. The low S/N spectra of 2S~0918-54,
XTE~0929-314 and A~1246-58 are difficult to classify, although
A~1246-58 shows some hints of detected C and O lines but not He
lines \citep{2008A&A...485..183I}.  Now \citet{icv05} suggest, based on
the type I X-ray bursts, that the donor of 2S~0918-54 is
more likely helium rich.  The optical spectrum of 4U~1916-05 does not
show strong He lines but is still best fitted with a He/N mixture.
Finally, the broadband UV spectrum of M15 X-2 is consistent with quite
strong emission lines of C and/or He \citep{2005ApJ...634L.105D}.

We therefore conclude that there is evidence for at least two helium
star or hybrid donors (4U~1626-67 and 4U~0614+09) but
more detailed observations are needed to classify the rest of 
the observed systems.

\subsection{Cataclysmic variables below the period minimum}\label{sec:evCV}

 Several population~I hydrogen-rich dwarf novae are known to
apparently have periods significantly below 80\,min  The SU UMa type
dwarf nova 1RXS J232953.9+062814 has an orbital period of 64.2\,min
\citep{2002ApJ...567L..49T,2002PASJ...54L..15U}.
\citet{2002ApJ...567L..49T} report for this system an ${\mathrm H\alpha}$
to He I $\lambda$6678 ratio  of 3.6, by at least factor 2 lower than
typical for SU UMa stars. According to \citet{2003ApJ...594..443G}
this system also has an anomalously high N\textsc{v}/C\textsc{iv} flux ratio. For V485 Cen,
\citet{1996A&A...311..889A} found an orbital period of 59\,min This
period is even shorter than the period minimum estimate for population~II
hydrogen-rich cataclysmic binaries \citep[70 min,
][]{1997A&A...320..136S}.  We may suspect that these stars belong to
the evolved CV family,  with an expected $X \approx 0.1$, $Y \approx
0.9$, see Fig.~\ref{fig:pre_min} but have to defer any firm
conclusions until sufficient observational data have been obtained.

\section{Discussion and Conclusions}\label{sec:discon}

We have computed the ranges of abundances (assuming initial Solar
  metallicity) in possible donor stars of ultra-compact binaries for
  the three different proposed formation channels, the white dwarf
  channel, the helium star channel and the evolved main-sequence star
  channel.

\rev{The \emph{main conclusion of our work} is that we have derived a
diagnostic for distinguishing the different channels (see
Figs.~\ref{fig:Nratios} and \ref{fig:Oratios})}.  First, the presence
of hydrogen unambiguously points to an evolved main-sequence donor and
rules out all other channels and vice versa.  Secondly, if no H is
detected, absence of He and N may point to a hybrid white dwarf.  Then
if N is detected, the N/C ratio is an effective discriminant between
helium white dwarf donors and helium star donors.  The difference
between He white dwarfs and He stars is that the former transfer
matter with equilibrium N/C which mildly depends on the mass of the
progenitor of the white dwarf (Fig.~\ref{fig:abund_HeWD1}) and is
always about $100$, while in the latter N/C has to be diminished
unless He didn't burn at all.  This is unlikely.  Once this
distinction is made, the N/O and N/He ratios can give further
information on the main-sequence progenitor mass for the helium white
dwarfs or the initial post-common-envelope period of the helium star
binary. If O or C is detected, the O/C ratio and the O/He ratio (or at
least their limits) are effective to distinguish hybrid white dwarf
from helium star donors. In the former, He has to be extinct and
$X_{\rm O}$ must exceed $X_{\rm C}$ by a large factor. In the latter,
$X_{\rm C}$, $X_{\rm O}$ and $X_{\rm He}$ vary within a broad range.

For the helium star channel the abundances are most
variable and depend on the amount of helium burnt before RLOF. The
least evolved donors have abundances similar to helium white dwarfs,
but, owing to higher masses of their main-sequence progenitors, they have
lower N/C ratios. In addition, even mild helium burning in the core
of a He-star progenitor immediately enhances the C abundance and brings
the N/C ratio down.  At the other extreme, there are similarities
between descendants of the initially most evolved helium stars and
hybrid white dwarfs which are dominated by C and O. However, as the
helium stars begin to contract before helium burning is completed,
even the most evolved helium star donors still have a mass
fraction of He $Y \approx 0.01$, in contrast to the completely
He-deficient hybrid white dwarf donors.

In the evolved main-sequence star channel, in the vast majority of
cases there is still some H left in the donors.  This distinguishes this
channel from the other formation channels.  The abundances of CNO-bicycle
species are close to the low-temperature burning equilibrium.  Only the
models of the most extreme evolutionary sequences, that evolve to very
short periods, get rid of all their hydrogen and may look like helium
white dwarf donors with low-mass main-sequence progenitors.

We have applied this scheme to the observed AM CVn systems and UCXBs and
conclude that, for a number of AM~CVn stars, there is evidence for helium
white dwarf donors (GP Com, CE~315 and SDSS~J0804+16). Neither H nor strong C
nor O lines are found and this argues against evolved main-sequence, hybrid
white dwarf or evolved helium star donors. For the UCXBs there are two
systems (4U~1626-67 and 4U~0614+09) in which detection of C and O
lines but no He lines suggests hybrid white dwarf or very evolved
helium star donors. For one UCXB (4U~1916-05), the detected He and N
lines suggest a He white dwarf or unevolved helium star donor.

 Another open question is the presence of $^{22}$Ne in the discs of
UCXBs.  \citet{jpc00} discovered an unusually high Ne/O ratio in the X-ray
spectrum of 4U~0614+091 and, based on similarity of the spectra,
claimed that several other X-ray sources are also UCXBs.
\citet{ynh02} have shown that the transfer of
Ne-enriched matter is possible in the late stages of evolution of UCXBs
with hybrid dwarf donors, thanks to uncovering the layers enriched in
Ne by gravitational sedimentation.  However, later observations showed
that the derived Ne/O ratio is variable and thus that the Ne may be
interstellar \citep{2005ApJ...627..926J}.

We conclude that more detailed observations in combination with
targeted searches for the ratios that we determined to be the best
diagnostics are a promising method for determining the relative
importance of the formation channels for ultra-compact binaries.

\medskip
\noindent
Detailed results of evolutionary computations for close binaries with
helium-star donors and white dwarf or neutron star accretors and plots
of the dependence of the masses of the donors, orbital periods of the
systems, mass-loss rates and abundances in the transferred matter on
the time passed since RLOF may be found at
\texttt{www.inasan.ru/$\sim$lry/HELIUM\_STARS/.}

\section*{Acknowledgments}

The authors thank A.V. Fedorova for providing unpublished details of her
evolutionary computations.  LRY acknowledges warm hospitality and support from
Department of Astrophysics of Nijmegen University.  GN is supported by
NWO-VENI grant 639.041.405 and NWO-VIDI grant 016.093.305.  LRY is supported
by RFBR grant 07-02-00454 and Presidium of the Russian Academy of Sciences
Program ``Origin, Evolution and Structure of the Universe Objects''.  MVvdS
acknowledges support from NSF CAREER Award AST-0449558 to Northwestern
University and from a CITA National Fellowship .  CAT thanks Churchill College
for a fellowship.  This research has made use of NASA's Astrophysics Data
System.

\bibliography{journals,Donor_Abundances,abundances} \bibliographystyle{mn2e}


\label{lastpage}

\end{document}